\documentclass{article}

\usepackage{amsfonts}
\usepackage{amsmath,amssymb,amscd,latexsym,euscript,mathrsfs}
\usepackage[all]{xy}

\usepackage{mathrsfs}

\usepackage{soul}

\DeclareMathOperator{\Lan}{{\rm Lan}}
\DeclareMathOperator{\Mor}{{\rm Mor}}
\DeclareMathOperator{\Ob}{{\rm Ob}}

\DeclareMathOperator{\Tran}{{\rm Tran}}

\newcommand{\AS}{{\mathcal AS}}
\newcommand{\ASb}{{\mathcal AS^{\flat}}}

\newcommand{\coLim}{\underrightarrow{\lim}}
\newcommand{\Lim}{\underleftarrow{\lim}}
\newcommand{\Set}{{\rm Set}}
\newcommand{\Mon}{{Mon}}

\newcommand{\NN}{{\,\mathbb N}}

\newcommand{\mC}{{\mathscr C}}
\newcommand{\mD}{{\mathscr D}}
\newcommand{\mE}{{\mathscr E}}
\newcommand{\mA}{{\mathcal A}}

\newcommand{\pt}{{\rm pt}}

\newcommand{\fC}{{\mathfrak{C}}}

\newcommand{\pright}{{\rightharpoonup}}

\newtheorem{theorem}{\bf{Theorem}}[section]
\newtheorem{lemma}[theorem]{\bf{Lemma}}
\newtheorem{proposition}[theorem]{\bf{Proposition}}
\newtheorem{corollary}[theorem]{\bf{Corollary}}
\newtheorem{definition}[theorem]{\bf{Definition}}
\newtheorem{example}[theorem]{\bf{Example}}

\def\geq{\geqslant}

\title
{Category of asynchronous systems and polygonal morphisms}

\author{ A. A. Husainov, husainov51@yandex.ru
}
\date{}

\begin{document}

\maketitle

\begin{abstract}
A {\em weak asynchronous system} is 
 a trace monoid with a partial action on a set. 
A {\em polygonal} morphism between weak asynchronous systems
 commutes with the actions and preserves the independence of events. 
We prove that the category 
of weak asynchronous systems and polygonal morphisms has 
all limits and colimits.
\end{abstract}

2010 Mathematics Subject Classification  18A35, 18A40, 18B20, 68Q85


\thispagestyle{empty}

\section*{Introduction}

Mathematical models of parallel systems find numerous applications 
in parallel programming.
They are applied for the development and verification of programs,
for searching deadlocks and estimation of runtime.
These models are widely applied to the description of semantics 
and the development of languages of parallel programming
\cite{mil1989}.

There are various models of parallel computing systems \cite{nie1991}. 
For example, for the solution of the dining philosophers problem,
 it is convenient to use higher dimensional automata \cite {faj1998}, 
but for a readers/writers problem, 
it is better to consider asynchronous systems \cite {X20042}.
For comparing the models, the adjoint functors between categories 
of these models are constructed 
  \cite{gou2002}, \cite{gou1995}, \cite{gou2012}, \cite{win1995}.

But, at comparision of asynchronous transition systems and 
higher dimensional automata, we face the open problem, 
whether there are colimits in the category of asynchronous systems.
We propose avoid this obstacle by constructing
 a cocomplete category of asynchronous systems, 
and it allows us to build adjoint functors in the standard way. 

The asynchronous system is a model of the computing system consisting of events 
(instructions, machine commands) and states. 
The states are defined by values of variables (or cells of memory).
Some events can occur simultaneously. 
The category of asynchronous systems for the first time has been studied 
by M. Bednarczyk \cite {bed1988}.  Class of morphisms was extended in \cite{bed2003}.

We consider asynchronous system as set with partial trace monoid action. 
We represent the action as total, adding to asynchronous system a
state ``at infinity''. 
Morphisms between trace monoids acting on the pointed sets 
lead to {\em polygonal} morphisms of weak asynchronous systems.

These morphisms have great value for studying homology groups 
of the asynchronous systems, introduced  in \cite {X20042}.
They also help in the studying homology groups of the Mazurkiewicz
trace languages and Petri nets \cite{X2012}. 
The review of the homology of asynchronous systems is contained in \cite {X20121}.

The paper consist of three sections. 
In the first, the category $FPCM$ of trace monoids and basic 
homomorhisms is investigated. It is proved, that in this category, 
there are limits (Theorem \ref{colf}) and colimits (Theorem \ref{cfpcm}) 
though even finite products do not coincide
with Cartesian products. The subcategory $FPCM^{\|} \subset FPCM $ 
with independence preserving morphisms
is studied. It is proved, that this subcategory is complete 
(Theorem \ref{colfpar}) and cocomplete (Theorem \ref {colfpcmpar}).
In the second section, the conditions of existence of limits and colimits 
in a category of diagrams over the fixed category are studied.
The third section is devoted to a category of 
weak asynchronous systems and polygonal морфизмов. 
Main results about completeness and cocompleteness of a category 
of weak asynchronous systems and polygonal morphisms are proved 
(Theorems \ref {asblim} and \ref {asbcolim}).


\section{Categories of trace monoids}

Bases of the trace monoid theory have been laid in \cite{car1969}. 
Applications in computer science belong to A. Mazurkiewicz \cite {maz1989}, 
V. Diekert, Y. M{\'e}tivier \cite {die1997}.
We shall consider a trace monoid category and basic homomorphisms
and its subcategory consisting of independence preserving homomorphisms.

The {\em diagram} is functor defined on a small category.
Our objective is research of a question on existence of limits 
and colimits of diagrams in these categories.

\subsection{Trace monoids}

A map $f: M\to M'$ between monoids is  {\em homomorphism},
if  $f(1)=1$  and  $f(\mu_1\mu_2)=f(\mu_1\mu_2)$ for all $\mu_1, \mu_2\in M$.
Denote by $Mon$ the category of all monoids and homomorphisms. 

Let $E$ be an arbitrary set. 
An {\em independence relation} on 
  $E$ is subset  $I\subseteq E\times E$ satisfying the following conditions:
\begin{itemize}
\item $(\forall a\in E)~ (a,a)\notin I$, 
\item $(\forall a, b\in E)~(a,b)\in I \Rightarrow (b,a)\in I$.
\end{itemize}

Elements $a,b\in E$ are  {\em independent}, if $(a,b)\in I$.

Let $E^*$ be the monoid of all  words $a_1 a_2\cdots a_n$ where
$a_1, a_2, \cdots, a_n\in E$ and $n\geq 0$,
 with operation of concatenation
$$
(a_1\cdots a_n)(b_1\cdots b_m)= a_1\cdots a_n b_1\cdots b_m\,.
$$
The identity $1$ is the empty word.

Let $I$ be an independence relation  on $E$. 
We define the equivalence relation  $ \equiv_I $ 
on  $E^*$ putting $w_1 \equiv_I w_2 $ if $w_2 $ can be receive 
from $w_1 $ by a finite sequence of adjacent independent elements.

\begin{figure}[h]
$$
\xymatrix{
a\ar@{-}[d] && b\ar@{-}[d]\\
e\ar@{-}[rr]\ar@{-}[rd] && c\ar@{-}[ld]\\
&d
}
$$
\caption{Independence graph}
\label{indgrph}
\end{figure}
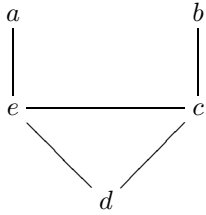

For example, for the set $E=\{a, b, c, d, e\}$ and  for the relation $I$ given by the 
adjacency graph drawn in Figure \ref{indgrph}
the sequence of permutations 
$$
	eadcc \stackrel{(e,a)}\to aedcc \stackrel{(e,d)}\to
adecc \stackrel{(e,c)}\to adcec \stackrel{(d,c)}\to acdec \stackrel{(e,c)}\to acdce
\stackrel{(d,c)}\to accde
$$
shows that $adecc\equiv_I accde$.

For every $w\in E^*$, its equivalence class $[w]$ 
 is called the {\em trace}.

\begin{definition}
Let $E$ be a set and let $I$ be an independence relation.
A trace monoid $M(E,I)$ is the set of equivalence classes 
$[w]$ of all $w\in E^*$ with the operation 
 $[w_1][w_2]=[w_1 w_2]$ for  $w_1, w_2\in E^*$.
\end{definition}
We emphasize that the set $ E $ can be infinite. 

In some cases, we omit the square brackets in the notations  
for elements of $M(E,I)$.
If $I=\emptyset$, then $M(E,I)$ is equal to the {\em free monoid} $E^*$.
If $I=((E\times E)\setminus \{(a,a)| a\in E\})$, then $M(E,I)$ is 
the {\em free commutative monoid}. In this case, we denote it by  
$M(E)$.

\subsection{The category of trace monoids and basic homomorphisms}

Let us introduce basic homomorphisms and we shall show, that the category 
of trace monoids and basic homomorphism is complete and cocomplete.

\begin{definition}
A homomorphism $f: M(E,I)\to M(E',I')$ is {\em basic}
if $f(E)\subseteq E'\cup \{1\}$.
\end{definition}

If $w= e_1\cdots e_n \in M(E,I)$ for some $e_1\in E$, ..., $e_n\in E$, 
then $n$ is called the {\em length} of the trace $w$.
It is easy to see, that a homomorphism will be basic, if and only if 
it does not increase length of elements of $M (E, I) $. 
Let $FPCM $ be a category of trace monoids and basic homomorphisms.

Consider the problem on existence of the products in $FPCM $.
The Cartesian product $M (E_1, I_1) \times M (E_2, I_2) $ 
will not be the product in $FPCM $. 
Thus for building products 
and other constructions, we shall consider partial 
maps as total, adding to them the element $*$.

Let $E_*= E\sqcup \{*\}$. 
We assign to each partial map 
$f: E_1 \pright E_2$, a total map $f_*: {E_1}_*\to {E_2}_*$ defined as
$$
f_*(a)= \left\{
\begin{array}{cc}
f(a), & \mbox{ if } f(a) \mbox{ defined},\\
 *, & \mbox{ otherwise.}
\end{array}
\right.
$$
 Any basic homomorphism $M(E_1, I_1)\to M(E_2,I_2)$
can be given by some partial map $f: E_1\pright E_2$.
We consider it as  the pointed total map
 $f_*: {E_1}_*\to {E_2}_*$ which brings any pair
 $(a,b)\in I_1\cup (E_1\times\{*\})\cup (\{*\}\times E_1)$ to the pair
$(f_*(a), f_*(b)) \in I_2\cup (E_2\times\{*\})\cup (\{*\}\times E_2)$.
It is clear that $f_*$ bring the elements of
 $\Delta_{{E_1}_*}= \{(a,a)| a\in {E_1}_*\}$
 to $(f_*(a),f_*(a))\in \Delta_{{E_2}_*}$.

Let $ComRel$ be the category of pairs 
$(E_*,T)$ where each pair consists of a pointed set $E_*$ and binary relation 
of {\em commutativity} $T\subseteq E_*\times E_*$  satisfying 
the following conditions
\begin{enumerate}
\item $(\forall a\in E_*)~ (a,*)\in T~ \&~ (*,a)\in T$ (commutativity with $1$),
\item $(\forall a\in E_*)~ (a,a)\in T$ (reflexivity),
\item $(\forall a, b\in E_*) (a,b)\in T \Rightarrow (b,a)\in T$ (symmetry).
\end{enumerate}
Morphisms $({E_1}_*, T_1) \stackrel{f}\to ({E_2}_*, T_2)$ in the category $ComRel$
are poinded maps $f: {E_1}_* \to {E_2}_*$
satisfying $(a_1, b_1)\in T_1 \Rightarrow (f(a_1),f(b_1))\in T_2$.

\begin{proposition}
The category $FPCM$ is isomorphic to $ComRel$.
\end{proposition}
{\sc Proof.} Define the functor $FPCM\to ComRel$ on objects by 
 $M(E,I)\mapsto (E_*, T)$
where  $T= I\cup (E\times\{*\})\cup (\{*\}\times E) \cup \Delta_{E_*}$.
The functor transforms basic homomorphisms 
$f: M(E_1,I_1)\to M(E_2,I_2)$ into the maps
 $f_*: {E_1}_* \to {E_2}_*$ assigning to
pairs $(a_1,b_1)\in T_1$ 
the pairs $(f_*(a_1),f_*(b_1))\in T_2$.

An inverse functor assigns to each object $ (E_*, T) $ of the category 
$ ComRel $ the trace monoid  $ M (E, I) $, where
\begin{equation}\label{ifort}
I = T\setminus \left( \{(a,a)| a\in E_*\} \cup \{(a,*)| a\in E\} \cup 
\{(*,a)| a\in E\}   \right),
\end{equation}
and to any morhism $({E_1}_*, T_1) \stackrel{f} \to ({E_2}_*, T_2)$ 
the homomorphism 
$\widetilde{f}: M(E_1,I_1)\to M(E_2, I_2)$
given at basic elements as $\widetilde{f}(e)= f(e)$ if $f(e)\in E_2$, 
and $\widetilde{f}(e)= 1$, if $f(e)=*$. 
\hfill $\Box$

Consider a family of trace monoids $\{M(E_{j},I_j)\}_{j\in J}$.
Transform it to family of pointed sets with commutativity
 relations $\{({E_j}_*, T_j)\}_{j\in J}$. 
The product of this family in the category $ComRel$
equals 
the Cartesian product $(\prod\limits_{j\in J}{E_j}_*, \prod\limits_{j\in J}T_j)$.
The category  $FPCM$ is isomorphic 
to  $ComRel$. Therefore, we obtain the following

\begin{proposition}\label{prodbas}
The category $FPCM$ 
has the products.
\end{proposition}

Any object $(E_*,T)$ of $ComRel$ corresponds to a trace monoid $M(E,I)$ 
with the set $E= E_*\setminus \{*\}$ and independence relation
defined by formula (\ref{ifort}).

It follows that the product of $M(E_j,I_j)$, $j\in J$  
has the set of generators 
$E= ( \prod\limits_{j\in J}E_{j*}) \setminus \{(*)\}$
where $(*)\in \prod\limits_{j\in J}E_{j*}$ 
denotes a family of elements each of which equals $*\in {E_j}_*$.
Let
$$
T_j = I_j \cup ( \{(a,a)| a\in {E_j}_*\} 
\cup \{(a,*)| a\in E_j\} \cup \{(*,a)| a\in E_j\}).
$$
The relation $I$ is received from
 $T= \prod\limits_{j\in J}T_j$ by the formula (\ref{ifort}).

\begin{example}
Let $J=\{1, 2\}$, $E_1=\{e_1\}, E_2= \{e_2\}$, $I_1= I_2= \emptyset$.
Then $M(E_1,I_1)\cong M(E_2, I_2)\cong \NN$ are isomorphic to the monoid generated 
by one element.
Compute $M(E,I)= M(E_1,I_1)\prod M(E_2,I_2)$. 
The set  $E_*$ equals ${E_1}_*\times {E_2}_*$.
In following picture at the left, it is shown the graph 
of the relation $T \subseteq E_* \times E_ * $ and on the right 
it is shown the graph of the relation $I$ obtained by the formula 
 (\ref {ifort}).
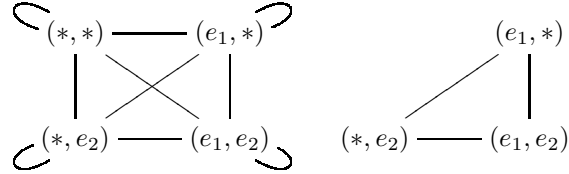
$$
\xymatrix{
		(*,*) \ar@{-}@(l,ul) \ar@{-}[r] \ar@{-}[rd] & (e_1, *) \ar@{-}@(ur,r) \ar@{-}[ld]\\
		(*, e_2) \ar@{-}@(l,dl) \ar@{-}[r] \ar@{-}[u] & (e_1, e_2) \ar@{-}@(r,dr) \ar@{-}[u]
}
\qquad
\xymatrix{
		 & (e_1, *)  \ar@{-}[ld]\\
		(*, e_2)  \ar@{-}[r]  & (e_1, e_2)  \ar@{-}[u]
}
$$
We see, that the product is isomorphic to a free commutative monoid generated 
by three elements.
\end{example}

\begin{theorem}\label{colf}
For each diagram  $D$ in  $FPCM$, there is the limit.
\end{theorem}
{\sc Proof.} Since $FPCM$ has all products, 
 it is enough existence of equalizers. 
Consider a pair  $\xymatrix{M(E_1,I_1)\ar@/^/[r]^f \ar@/_/[r]_g & M(E_2,I_2)}$
of basic homomorphisms.
Let $E=\{e\in E_1 ~|~ f(e)=g(e)\}$. 
The submonoid of $M(E_1,I_1)$ generated by $E$ is a trace monoid
$M(E,I)$ with the independence relation $I=I_1 \cap (E\times E)$.
Consider an arbitrary basic homomorphismй $h': M(E',I')\to M(E_1, I_1)$
such that $g(h'(e'))=f(h'(e'))$ for all $e'\in E'$.
Obtain $h'(e')\in E\cup \{1\}$. It follows that $h'$ maps $M(E',I')$ into $M(E,I)$
and the following triangle is commutative:
$$
\xymatrix{
M(E,I) \ar[rr]^{\subseteq} & & M(E_1, I_1)\\
& M(E',I') \ar[lu] \ar[ru]_{h'}
}
$$
Therefore, the inclusion $M(E,I)$ into $M(E_1,I_1)$ is equalizer
of the pair $(f,g)$.
\hfill $\Box$

\begin{proposition}\label{rightadj}
Let $\Ob\Mon \to \Ob FPCM$ be the map carried each monoid $M$
to a trace monoid $M(M\setminus \{1\}, I_M)$ with
$$
I_M = \{(\mu_1,\mu_2)\in (M\setminus \{1\})\times (M\setminus \{1\})
 | ~ \mu_1\not=\mu_2 ~\& ~\mu_1\mu_2= \mu_2\mu_1\}.
$$
This map can be extended to a functor $R: \Mon\to FPCM$ right adjoint 
to the inclusion $U: FPCM\to \Mon$.
\end{proposition}
{\sc Proof.} Define a homomorphism $\varepsilon_M: M(M\setminus \{1\}, I_M)\to M$
setting $\varepsilon(\mu)=\mu$ on the generators of 
  $M(M\setminus \{1\}, I_M)\to M$.
It easy to see that for each homomorphism  $f: M(E,I)\to M$, there exists
unique basic homomorphism  $\overline{f}$ making the following 
diagram commutative
$$
\xymatrix{
M(M\setminus \{1\}, I_M) \ar[rr]^{\varepsilon_M} && M\\
& M(E,I)\ar@{-->}[lu]^{\overline{f}} \ar[ru]_f
}
$$
It is defined by $\overline{f}(e)=f(e)$ on elements $e\in E$.
This homomorphism is couniversal arrow.
By the universal property,
the map
$M\mapsto (M(M\setminus \{1\}, I_M),\varepsilon_M)$
uniquely extends up to the right adjoint functor.
\hfill
$\Box$

\begin{theorem}\label{cfpcm}
The category $FPCM$ is cocomplete and the inclusion functor 
 $FPCM$ into the category $\Mon$ preserves all colimits.
\end{theorem}
{\sc Proof.} Let $D: J\to FPCM$ be a diagram
with values $D(j)=M(E_j,I_j)$.
Consider   $\coLim^{J}D$  in the category $Mon$ of all monoids.
The colimit is isomorphic to a quotient monoid 
 $\coprod_{j\in J} M(E_j, I_j)/ \equiv$ obtained from 
the coproduct in $Mon$ by
identifications of elements
$e_j \equiv D(j\to k)e_j$.
It follows that the colimit is generated by the disjoint union $\coprod\limits_{j\in J}E_j$ 
and represented by the following equations:
\begin{enumerate}
\item for all $ j\in J$,  $ (e, e')\in I_j$ it is true $ee'\equiv e'e$, \label{eq1}
\item  if $e'_k= D(j\to k)(e_j)$ for some $e_j\in E_j$, $e'_k\in E_k$, 
then $e_j\equiv e'_k$, \label{eq2}
\item $e_j\equiv 1$ if $M(j\to k)(e_j)=1$.
\end{enumerate}
This monoid is generated by a set $E $
 received of a quotient set of $\coprod \limits _ {j \in J} E_j $ 
under the equivalence relation containing pairs type (\ref{eq2})
by removing the classes containing elements $e_j \equiv 1 $.
The equations (\ref{eq1}) give the relation $I$.
We obtain the trace monoid $\coLim^{J}D=M(E,I)$.
The morphisms of colimiting cone are basic homomorhisms
sending to every
$e_j$ its equivalence class or  $1$.
For any other cone $f_j: M(E_j, I_j)\to M(E',I')$
consisting of basic homomorphisms,   
the morphism 
 $\coLim^{J}D\to M(E',I')$
assigns to each class $[e_j]$ the element $f_j(e_j)$.
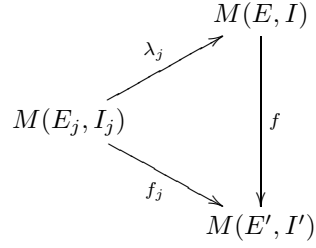
$$
\xymatrix{
	& M(E,I) \ar[dd]^{f}\\
  M(E_j, I_j)\ar[ru]^{\lambda_j} \ar[rd]_{f_j} \\
	& M(E',I')
}
$$
Therefore, $FPCM$ has all colimits.

It follows from \ref{rightadj} that the inclusion 
$FPCM \subset Mon$ preserves all colimits 
as having right adjoint \cite{mac1998}.
\hfill
$\Box$

\begin{example}\label{excoeq}
Consider the free commutative monoid $M(\{a,b\})$ 
and the trace monoid $M= M(\{c, d, e\}, \{(c,d), (d,c), (d,e), (e,d)\})$.
Let $f, g: M\{a,b\}\to M$ be two homomorphisms defined as $f(a)=c$, 
$g(b)=d$,  $g(a)=d$, $g(b)=c$.
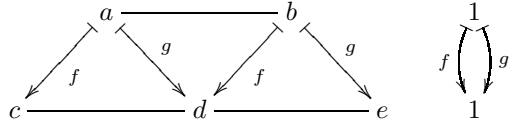
$$
\xymatrix{
& a \ar@{|->}[ld]^f \ar@{|->}[rd]^g \ar@{-}[rr] & & b\ar@{|->}[ld]^f \ar@{|->}[rd]^g & & 1 
\ar@{|->}@/_/[d]_f \ar@{|->}@/^/[d]^g\\
c \ar@{-}[rr] && d \ar@{-}[rr] & & e 
& 1
}
$$
The coequalizer of $f,g$ is the trace monoid generated by $c, d, e$
with equations  $c=a=d=b$,
 $cd=dc$, $de=ed$.
In the picture in the top line,  it is shown the independence 
relation for $M(\{a, b\})$, and in bottom for $M$.
Consequently, the coequalizer is equal to the free commutative monoid 
generated by one element.
\end{example}

\subsection{Independence preserving basic homomorphisms}

We prove that the category of trace monoids 
and  independence preserving homomorphisms
has all limits and colimits.

\begin{definition}\label{defind}
A basic homomorphism $f: M(E,I)\to M(E',I')$ is called 
{\em independence preserving} if
for all $a,b\in E$, the following implication is carried out
$$
 (a,b)\in I \Rightarrow (f(a)\not= f(b))~ \vee~ (f(a)=f(b)=1)~.
$$
\end{definition}

It is easy to see, that this implication is equivalent to the condition
$$
 (a,b)\in I \Rightarrow (f(a), f(b))\in I'~ \vee~ f(a)=1 ~\vee~ f(b)=1~.
$$

It follows that the class of independence preserving homomorphisms
is closed under composition.
Let $FPCM^{\|} \subset FPCM$ be the subcategory consisting of all
trace monoids and independence preserving  basic homomorphisms.

Let us prove the existence of the products in the $ FPCM^{||} $.
For this purpose we introduce the following  {\em partial independence relation}.

\begin{definition}
Let $E$ be a set. A {\em partial independence relation  on $E$}
is a subset $R\subseteq E_*\times E_*$ satisfying the followng conditions:
\begin{enumerate}
\item $(\forall a\in E_*)~ (a,*)\in R ~\&~ (*,a)\in R$;
\item $(\forall a\in E_*)~ (a,a)\in R \Rightarrow a=*$;
\item $(\forall a, b\in E_*)~ (a,b)\in R \Leftrightarrow  (b,a)\in R$.
\end{enumerate}
\end{definition}

Let $IndRel$ be the category of pairs $(E_*,R)$ consisting 
of pointed sets $E_*$ and partial independence relations  
$R\subseteq E_*\times E_* $.
Its morphisms $(E_*,R) \stackrel{f}\to (E_*',R')$ defined as 
pointed maps $f: E_*\to E_*'$ satisfying the following conditions:
$$
(a,b)\in R \Rightarrow (f(a), f(b))\in R'.
$$

\begin{proposition}
The category $FPCM^{||}$ is isomorphic to $IndRel$.
\end{proposition}
{\sc Proof.} Define the functor  $FPCM^{||}\to IndRel$ as
sending  
$M(E,I)$ to $(E_*,R)$ where $R= I\cup (E_*\times \{*\})\cup (\{*\}\times E_*)$.
The inverse functor $IndRel \to IndRel$ carries any object 
$(E_*,R)$ to the monoid $M(E,I)$ where 
$I= R \setminus (E_*\times \{*\})\cup (\{*\}\times E_*)$. 
This functor send morphisms of $IndRel$ to  
independence preserving morphisms.
\hfill $\Box$

\begin{corollary}
The category $FPCM^{||}$ has all products.
\end{corollary}

Moreover, it is true the following

\begin{theorem}\label{colfpar}
The category $FPCM^{\|}$ has all limits. 
The inclusion functor $FPCM^{\|}\subset FPCM$
 preserves equalizers.
\end{theorem}
{\sc Proof.} Since $FPCM^{\|}$ has products,
it it enough to prove the existence equalizers.
For any pair of basic homorphisms 
$\xymatrix{M(E_1,I_1)\ar@/^/[r]^f \ar@/_/[r]_g & M(E_2,I_2)}$
in the category $FPCM$ its equalizer is the inclusion  $M(E,I)\subseteq M(E_1,I_1)$,
where $E=\{e\in E_1| f(e)=g(e)\}$ and $I= I_1\cap (E\times E)$.
Inclusion preserves independence.
Consider a trace monoid 
 $M(E',I')$ with a independence preserving homomorphism
$h: M(E',I')\to M(E_1,I_1)$ satisfying $fh=gh$.
Since $h(E')\subseteq E$, there is a basic homomorphism $k$ drawn 
by dashed arrow in the diagram:
$$
\xymatrix{
M(E,I) \ar[r]^{\subseteq} & M(E_1,I_1) \ar@/^/[r]^f \ar@/_/[r]_g & M(E_2,I_2)\\
M(E',I') \ar[ru]_{h} \ar@{-->}[u]^k
}
$$
We have  $k(e')=h(e')$ for all $e'\in E'$. The homomorphism $h$ preserves 
independence. Hence, for all
$(a',b')\in I'$, the condition $k(a')=1 \vee k(b')=1 \vee (k(a'),k(b'))\in I_1$
holds.
Thus, $k$ preserves independence. 
Equalizers is constructed in the category $FPCM$. Therefore the 
inclusion $FPCM^{\|}\subset FPCM$ preserves equivalizers.
\hfill $\Box$

We now turn to the colimit.

\begin{theorem}\label{colfpcmpar}
The category $FPCM^{\|}$ is cocomplete.
\end{theorem}
{\sc Proof.} The coproduct of trace monoids $\{M(E_i, I_i)\}_{i\in J}$ is 
a monoid given by generators  $\coprod\limits_{i\in J}E_i$ and relations 
 $ab=ba$ for all  $(a,b)\in \coprod\limits_{i\in J}I_i$.
It is easy to see that it is coproduct in the category $FPCM^{||}$.
Hence, it is sufficient to prove the existence coequalizers.
For this purpose, consider an arbitrary pair of morphisms
$\xymatrix{M(E_1,I_1) \ar@/^/[r]^{f} \ar@/_/[r]_{g} & M(E_1,I_1)}$
in the category $FPCM^{\|}$.
Let $h: M(E_2, I_2)\to M(E,I)$ be the coequalizer in the 
category $FPCM$. For each $h': M(E_2,I_2)\to M(E',I')$, there exists
a unique $k$ making commutative triangle in the following diagram 
$$
\xymatrix{
	M(E_1,I_1) \ar@/^/[r]^{f} \ar@/_/[r]_{g}
		 & M(E_1,I_1)\ar[d]_{h'} \ar[r]^h & M(E,I) \ar@{-->}[ld]_{\exists!k}\\
& M(E',I')
}
$$
If $h'$ preserves independence, then the following implication is true:
\begin{equation}\label{pres}
(\forall (a,b)\in I_2) (h'(a)=h'(b) \Rightarrow h'(a)=1 ~\&~ h'(b)=1).
\end{equation}

Let $\equiv_h$ be the smallest congruence relation for which 
$h(a)\equiv_h 1$ and 
$h(b)\equiv_h 1$ if $(a,b)\in I_2$ satisfies $h(a)=h(b)$.
Denote by $cls: M(E,I)\to M(E,I)/\equiv_h$ the homomorphism
assigning to any $e\in E_*$ its class $cls(e)$ of the congruence. 
If $h'$ preserves independence, then it follows from (\ref{pres}) and 
$kh=h'$ that
$$
(\forall (a,b)\in I_2) k(h(a))=k(h(b)) \Rightarrow k(h(a))=1 ~\&~ k(h(b))=1.
$$
We see that $k$ has constant values on each congruence class $cls(e)$ 
where $e\in E_*$. Hence, we can define
 a map $k': M(E,I)/\equiv_h \to M(E',I')$ by $k'(cls(e))=k(e)$ for all $e\in E_*$.
The homomorphism $k'$ is unique for which $k'\circ cls\circ h= h'$.
Therefore,  $cls\circ h: M(E_2,I_2)\to M(E,I)/\equiv_h$ is the coequalizer 
of $(f,g)$.
\hfill $\Box$

In Example \ref{excoeq}, we have $h(c)=h(d)=h(e)$.
Since $(c,d)\in I_2$ and $(d,e)\in I_2$, we have 
$cls\circ h(c)=1$, $cls\circ h(d)=1$, $cls\circ h(e)=1$.
Therefore, the coequalizer equals $\{1\}$.  

\section{Category of diagrams with various domains}

This Section is auxiliary also does not contain new results.
A {\em diagram in a category $\mA$} is a functor 
$ \mC \to \mA $ defined on some small category $\mC$.
We shall consider categories of the diagrams accepting  
values in some fixed category. 
Let us study the conditions providing completeness
 or cocompleteness of this category.

\subsection{Morphisms and objects in a digram category}

Let $\mA$ be a category and let $F: \mC\to \mA$ be a diagram.
Denote this diagram by $(\mC, F)$ specifying its domain $\mC$.

Let $(\mC,F)$ and $(\mD, G)$ be diagrams in $\mA$.
A {\em morphism of the diagrams} $(\Phi, \xi): (\mC, F)\to (\mD, G)$ is given 
by a pair $(\Phi, \xi)$ consisting of a functor
$\Phi: \mC\to \mD$ and natural transformation $\xi: F\to G\Phi$

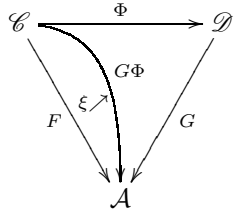
$$
\xymatrix{
	\mC \ar[rr]^{\Phi} \ar[rdd]_F^{\xi\nearrow} \ar@(r,u)[rdd]^{G\Phi} && \mD\ar[ldd]^G\\
\\
	 & \mA
}
$$

Define  the
{\em identity morphism} by the formula  
 $1_{(\mC,F)}= (1_{\mC}, 1_{F})$ where $1_{\mC}: \mC\to \mC$ is the identity 
functor and $1_F: F\to F$ is the identity natural transformation. 
The {\em composition of morphisms} 
$$
(\mC,F)\stackrel{(\Phi,\xi)}\to (\mD, G) \stackrel{(\Psi,\eta)}\to (\mE, H)
$$
is defined as a pair $(\Psi\Phi, (\eta*\Phi)\cdot\xi)$
where $\eta*\Phi: G\Phi\to H\Psi\Phi$ is a natural transformation given 
by a family of morphisms specified as a family of morphisms
$$
	(\eta*\Phi)_c = \eta_{\Phi(c)}: 
		G(\Phi(c))\to H(\Psi(\Phi(c))), ~ c\in \Ob\mC,
$$
and $(\eta*\Phi)\cdot\xi$ is the composition of natural 
transformations 
$F\stackrel{\xi}\to G\Phi \stackrel{\eta*\Phi}\to H\Psi\Phi$.
The composition is associative.

Let $Cat$ be the category of small categories and functors.
Denote by $(Cat, \mA)$ the category of diagrams in $\mA$ 
and morphisms of diagrams.

For any subcategory ${\mathfrak C} \subseteq Cat$, 
we consider diagrams $F: \mC\to \mA$ defined on categories $\mC \in {\mathfrak C}$.
Such diagrams with morphisms $(\Phi,\xi): (\mC, F)\to (\mD,G)$ where 
$\Phi\in \Mor{\mathfrak C}$, will be make a subcategory 
of  $(Cat, \mA)$. 
Denote this subcategory by $({\mathfrak C}, \mA)$.

\subsection{Limits in a category of diagram}

Let $J$ be a small category. 
In some cases, the diagrams are conveniently denoted, specifying their values 
on objects.
For example, we will denote  by $\{A_i\}_{i\in J}$ the diagram $J\to \mA$ with values
$A_i$ on objects $i\in J$ and  $A_{\alpha}: A_i\to A_j$ on morphisms 
$\alpha: i\to j$ of $J$.
We say that a category  $\mA$ has $J$-limits if 
every diagram  $\{A_i\}_{i\in J}$ in $\mA$ has a limit.
If $\mA$ has $J$-limits for all small categories $J$, 
then $\mA$ is said to be a {\em complete category} or a {\em category with all limits}.

We will consider subcategories ${\mathfrak C}\subseteq Cat$ with $J$-limits.
But the  $J$-limits in ${\mathfrak C}$  need not be isomorphic 
to the $J$-limits in $Cat$.

\begin{proposition}\label{limdg}
Let $\mA$ be a complete category and let $J$ be a small category.
If a subcategory ${\mathfrak C}\subseteq Cat$ has $J$-limits, then 
the category $({\mathfrak C}, \mA)$ has $J$-limits.
In particular, if ${\mathfrak C}\subseteq Cat$ is a complete category,
then the category $({\mathfrak C}, \mA)$ is complete.
\end{proposition}
{\sc Proof.}
Let $\{(\mC_i, F_i)\}_{i\in J}$ be a diagram in $({\mathfrak C}, \mA)$.
One given by a diagram  $\{\mC_i\}_{i\in J}$ with functors $\mC_{\alpha}: \mC_i\to \mC_j$
and natural transformations $\varphi_{\alpha}: F_i\to F_j\mC_{\alpha}$.
Let $p_i: \Lim_J\{\mC_i\}\to \mC_i$ is the limit cone of the 
diagram  $\{\mC_i\}_{i\in J}$ in ${\mathfrak C}$.
The compositions $F_i\circ p_i$ belong to the category $\mA^{\Lim_J\{\mC_i\}}$. 
The natural transformations 
$F_i p_i \stackrel{\varphi_{\alpha}*p_i}\to F_j\mC_{\alpha}p_i \stackrel{=}\to F_j p_j$ 
give the functor  $J\to \mA^{\Lim_J\{\mC_i\}}$. 
Let $\Lim_{J}\{F_i p_i\}\in \mA^{\Lim_J\{\mC_i\}}$ be its limit.
Denote by $\pi_i: \Lim_J\{F_i p_i\}\to F_i p_i$ the limit cone.
It easy to see that morphisms  
$(p_i, \pi_i): (\Lim_J\{\mC_i\}, \Lim_J\{F_i p_i\})\to (\mC_i,F_i)$ of diagrams
make the cone over the diagram in  $({\mathfrak C}, \mA)$. 
Considering an another cone $(r_i,\xi_i): (\mC, F)\to (\mC_i, F_i)$
it can be seen that there exists the unique morphism
 $(r,\xi)$ making the commutative triangle
$$
\xymatrix{
(\mC, F) \ar[rr]^{(r_i,\xi_i)} \ar@{-->}[rd]_{(r,\xi)} && (\mC_i, F_i)\\
& (\Lim_J\{\mC_i\}, \Lim_J\{F_i p_i\}) \ar[ru]_{(p_i,\pi_i)}
}
$$
It follows that the limit is isomorphic to $(\Lim_J\{\mC_i\}, \Lim_J\{F_i p_i\})$.
\hfill $\Box$

\subsection{Colimits in a category of diagrams}

Let ${\mathfrak C}\subseteq Cat$ be a subcategory. 
Consider an arbitrary category $\mA$.
We shall prove that if the colimits exist in ${\mathfrak C}$, 
then those exist in $({\mathfrak C}, \mA)$.
For any functor $\Phi: \mC\to \mD$, we denote by 
$\Lan^{\Phi}: \mA^{\mC}\to \mA^{\mD}$ the left Kan extension functor \cite{mac1998}.
Its properties are well described in \cite{mac1998}. 
This functor is characterized as a left adjoint 
to the functor $ \Phi^*: \ mA^{\mD} \to \mA^{\mC} $
assigning to each diagram $ F: \mD \to \mA $ 
the composition $ F \circ \Phi $, and to the natural transformation 
$ \eta: F \to G $ the natural transformation $ \eta* \Phi $.

\begin{proposition}\label{excolpol}
Let  ${\mathfrak C}\subseteq Cat$ be a category with all colimits.
Then, for any cocomplete category $\mA$, the category
 $({\mathfrak C}, \mA)$ has all colimits.
\end{proposition}
{\sc Proof.} Consider a diagram  $\{(\mC_i, F_i)\}_{i\in J}$ in $({\mathfrak C}, \mA)$.
As above, each morphism $ \alpha: i \to j $ is mapped to the natural transformation
$\varphi_{\alpha}: F_i\to F_j\mC_{\alpha}$. 
Let $\coLim^{J}\{\mC_i\}$ be the colimit of the diagram in ${\mathfrak C}$. 
Denote by $q_i: \mC_i \to \coLim^J\{\mC_i\}$ morphisms of the colimit cone.
Consider the Kan extensions 
$\Lan^{q_i}F_i$ and the units of adjunction 
$$
\xymatrix{
\mC_i \ar[rr]^{q_i} \ar[rd]_{F_i}^{\nearrow\eta_i} && \coLim^J\{\mC_i\}\ar[ld]^{\Lan^{q_i}F_i}\\
& \mA
}
$$
We get the diagram in the category $\mA^{\coLim^J\{\mC_i\}}$
consisting of objects $\Lan^{q_i}F_i$
and morphisms given at $\alpha: i\to j$ by the compositions 
$$
\Lan^{q_i}F_i \stackrel{\Lan^{q_i}(\varphi_{\alpha})}\to 
\Lan^{q_i}F_j\mC_{\alpha} \stackrel{=}\to \Lan^{q_j}\Lan^{\mC_{\alpha}}F_j\mC_{\alpha} 
\stackrel{\Lan^{q_j}(\varepsilon_{\alpha})}\to \Lan^{q_j}F_j 
$$
where $\varepsilon_{\alpha}: \Lan^{\mC_{\alpha}}(F_j\mC_{\alpha}) 
\to F_j$ are counits of adjunction.
Let $\coLim^J\{\Lan^{q_i}F_i\}$ be the colimit of this diagram. 

Prove that 
$(\coLim^J\{\mC_i\}, \coLim^J\{\Lan^{q_i}F_i\})$ is a colimit 
of the diagram $\{(\mC_i, F_i)\}_{i\in J}$
in $({\mathfrak C}, \mA)$.
For this purpose, consider an arbitrary 
(direct) cone  $(\mC_i, F_i) \to (\mC,F)$ 
over $\{(\mC_i, F_i)\}_{i\in J}$ in the category $({\mathfrak C}, \mA)$.
One is given by some functors 
$$
\xymatrix{
\mC_i \ar[rr]^{r_i} \ar[rd]_{F_i}^{\nearrow\psi_i} && \mC\ar[ld]^{F}\\
& \mA
}
$$
and natural transformations $\psi_i: F_i\to Fr_i$
for which the following diagrams are commutative
$$
\xymatrix{
(\mC_i, F_i)\ar[d]_{(\mC_{\alpha}, \varphi_{\alpha})} \ar[r]^{(r_i,\psi_i)} & (\mC,F)\\
(\mC_j, F_j) \ar[ru]_{(r_j, \psi_j)}
}
\quad
\xymatrix{
F_i \ar[r]^{\psi_i}\ar[d]^{\varphi_{\alpha}} & Fr_i\\
F_j\mC_{\alpha}\ar[r]_{\psi_j*\mC_{\alpha}} & Fr_j\mC_{\alpha}\ar@{=}[u]
}
$$
Since $\coLim^J\{\mC_i\}$ is the colimit in  ${\mathfrak C}$, the unique 
functor $r: \coLim^J\{\mC_i\}\to \mC$ is corresponded to the functors of 
this cone 
$r_i: \mC_i\to \mC$, such that 
 $r_i=rq_i$ dor all $i\in J$ where $q_i: \mC_i\to \coLim^J\{\mC_i\}$ is the 
colimit cone.

For any $i\in J$, the functor $\Lan^{q_i}$ is left adjoint to $q_i^*$.
Hence, there exists a bijection between natural transformations
$$
	F_i \stackrel{\psi_i}\to Fr_i=Frq_i \quad \mbox{ and } \quad \Lan^{q_i}F_i 
\stackrel{\overline{\psi_i}}\to Fr\,.
$$
This bijection maps each commutative triangle in $\mA^{\mC_i}$  to
the commutative triangle in  $\mA^{\coLim^J\{\mC_i\}_{i\in J}}$:
$$
\xymatrix{
F_i \ar[r]^{\psi_i}\ar[d]^{\varphi_{\alpha}} & Frq_i\\
F_j\mC_{\alpha}\ar[r]_{\psi_j*\mC_{\alpha}} & Frq_j\mC_{\alpha}\ar@{=}[u]
}
\quad \mapsto \quad 
\xymatrix{
\Lan^{q_i}F_i \ar[d] \ar[r]^{\overline{\psi_i}} & Fr\\
\Lan^{q_j}F_j \ar[ru]_{\overline{\psi_j}}
}
$$

For the diagram 
$$
\xymatrix{
F_i \ar[rr]^{\psi_i}\ar[d]^{\varphi_{\alpha}} && Frq_i= Frq_j\mC_{\alpha}\\
F_j\mC_{\alpha}\ar[rru]_{\psi_j*\mC_{\alpha}}
}
$$
we have the commutative diagram in $\mA^{\mC_j}$
$$
\xymatrix{
\Lan^{\mC_{\alpha}}F_i \ar[d]_{\overline{\varphi_{\alpha}}} 
 \ar[rr]^{\overline{\psi_i}} && Frq_j\\
F_j \ar[rru]_{\psi_j}
}
$$
Applying $\Lan^{q_j}$, we obtain the commutative diagram
$$
\xymatrix{
\Lan^{q_j}\Lan^{\mC_{\alpha}}F_i \ar[d]_{\Lan^{q_j}{\overline{\varphi_{\alpha}}}} \ar[rr]^{\Lan^{q_j}{\overline{\psi_i}}} 
&& \Lan^{q_j}Frq_j \ar[r]^{{\varepsilon_j}_{Fr}}& Fr\\
\Lan^{q_j}F_j \ar[rru]_{\Lan^{q_j}({\psi_j})}
}
$$
which leads us to the (direct) cone over the diagram $\{\Lan^{q_i}F_i\}_{i\in J}$
$$
\xymatrix{
\Lan^{q_i}F_i \ar[d] \ar[r] & Fr\\
\Lan^{q_j}F_j \ar[ru]
}
$$
This cone gives the morphism
$\coLim^J\{\Lan^{q_i}F_i\}\to Fr$ in  $\mA^{\coLim^J\{\mC_i\}}$
which define a unique morphism in $({\mathfrak C}, \mA)$
making commutative triangles
$$
\xymatrix{
(\coLim^J\{\mC_i\}, \coLim^J\{\Lan^{q_i}F_i\})\ar@{-->}[rr] && (\mC,F)\\
& (\mC_i, F_i)\ar[lu] \ar[ru]
}
$$
Therefore, the diagram $(\coLim^J\{\mC_i\}, \coLim^J\{\Lan^{q_i}F_i\})$ in 
 $\mA$ 
is
the colimit of the diagram  $\{(\mC_i, F_i)\}_{i\in J}$ 
in the category $({\mathfrak C}, \mA)$.
\hfill $\Box$

\section{Category of pointed polygons on trace monoids}

We apply auxiliary propositions from section 2 to categories 
$\mA= \Set_*$, $\fC= FPCM$ and $\fC=FPCM^{\|}$.
Then we shall establish communications between a category of asynchronous
 systems and categories of right $M(E, I)$-sets and 
we investigate a category of asynchronous systems and polygonal morphisms.

\subsection{Category of state spaces}

A {\em state space} $(M(E,I),X)$
consists of a trace monoid $M(E,I)$ with an  action on a pointed set $X$
by some operation
 $\cdot: X\times M(E,I)\to X$, 
$x\mapsto x\cdot w$ for $x\in X$, $w\in M(E,I)$.
Since the monoid is a category with a unique object, 
we can consider the state space as a functor
$X: M(E,I)^{op}\to Set_*$ sending the unique object to
the pointed set $X$ and morphisms 
 $w\in M(E,I)$ to maps $X(w): X\to X$ given as $X(w)(x)=x\cdot w$.
Here we denote by $X$ the pointed set on which the monoid acts 
as well as functor defined by this action.

\begin{definition}
A {\em morphism of state spaces}
$$
(M(E,I),X)\to (M(E',I'),X')
$$
is a pair $(\eta, \sigma)$ where 
$\eta: M(E,I)\to M(E',I')$ is a basic homomorphism and 
$\sigma: X\to X'\circ\eta^{op}$ is a natural transformation.
\end{definition}

A morphism of state spaces is possible to represent by means 
of the diagram

$$
\xymatrix{
M(E,I)^{op}\ar[rdd]_{X}^{\sigma\nearrow} \ar[rr]^{\eta^{op}} && M(E',I')^{op}\ar[ldd]^{X'}\\
\\
& Set_*
}
$$

The category of state space is isomorphic to $(FPCM,Set_*)$.

By Proposition \ref{limdg}, if a subcategory ${\mathfrak C}\subseteq Cat$
has $J$-limits, then $({\mathfrak C}, \Set_*)$ has  $J$-limits. 
For $\fC= FPCM$ and for discrete category $J$ with $\Ob(J)= \{1,2\}$,
it follows from Proposition \ref{limdg}, the following

\begin{proposition}\label{prodinss}
Let $(M(E_1, I_1),X_1)$
and $(M(E_2, I_2),X_2)$ be state spaces.
Their product in  $(FPCM, Set_*)$ is a state space
$$
(M(E_1, I_1)\prod M(E_2, I_2), X_1\circ \pi_1^{op}\times X_2\circ \pi_2^{op})
$$
where $\pi_i: M(E_1, I_1)\prod M(E_2, I_2) \to M(E_i, I_i)$ are the projections 
of the product in the category $FPCM$ for $i\in \{1,2\}$.
\end{proposition}

\begin{definition}
A morphism  $(\eta, \sigma): (M(E,I),X)\to (M(E',I'),X')$ of state spaces
is independence preserving if 
$\eta: M(E,I)\to M(E',I')$ is  independence preserving.
\end{definition}

Let $(FPCM^{\|}, Set_*)\subset (FPCM, Set_*)$ be the subcategory 
of all state spaces and independence preserving morphisms.

\begin{proposition}\label{cpolpar}
The categories  $(FPCM, Set_*)$ and $(FPCM^{\|}, Set_*)$ are complete.
\end{proposition}
{\sc Proof.} The category $FPCM$ is complete by Theorem \ref{colf}
and $FPCM^{\|}$ is complete by Theorem \ref{colfpar}. 
Proposition \ref{limdg} gives completeness of $(FPCM, Set_*)$ and $(FPCM^{\|}, Set_*)$.

\begin{proposition}\label{cocpolpar}
The categories $(FPCM,Set_*)$ and $(FPCM^{\|},Set_*)$ are cocomplete.
\end{proposition}
{\sc Proof.}
First statement follows from Theorem  \ref{cfpcm} 
and  Proposition \ref{excolpol} applied to
  ${\mathfrak C}= FPCM$ and $\mA = Set_*$. The second statement follows 
from Theorem
\ref{colfpcmpar} 
and Proposition \ref{excolpol}.
\hfill $\Box$.

\subsection{Category of weak asynchronous system and polygonal morphisms}

\begin{definition}\label{defasyns}
{\em The weak asynchronous system} $ \mA= (S, s_0, E, I, \Tran) $ 
consist of a set $S $ which elements called {\em states}, 
{\em an initial state} $s_0 \in S_* $, a set $E $ of {\em events}, 
the irreflective symmetric relation $I \subseteq E \times E $ of {\em independence}, 
satisfying the conditions
\begin{itemize}
\item  If $(s,a,s')\in \Tran$ $\&$ $(s,a,s'')\in \Tran$, then $s'=s''$.
\item  If $(a,b)\in I ~ \& ~(s, a, s')\in \Tran ~\& ~
(s', b, s'')\in \Tran$, then there exists $s_1\in S$ such that $(s,b,s_1)\in \Tran$
$\&$ $(s_1, a, s'')\in \Tran$. 
\end{itemize}
\end{definition}

If we add to Definition \ref{defasyns} the conditions $s_0\in S$ and 
 $S\not= \emptyset$, then we obtain
asynchronous systems in the sense of M. Bednarczyk
\cite{bed1988}. If more than that, we require the condition
$(\forall e\in E)(\exists e, e'\in S)~ (s,e,s')\in \Tran$, 
 then we get an {\em asynchronous transition system} 
 \cite{win1995}.

\begin{lemma}
Every weak asynchronous system  $(S, s_0, E, I, \Tran)$ gives
a state space  $(M(E,I), S_*)$ with a distinguished element $s_0\in S_*$
wherein the action is defined by 
$$
(s, [e_1\cdots e_n])\mapsto (\ldots((s\cdot e_1)\cdot e_2) \ldots \cdot e_n),
$$
for all $s\in S_*$ and $e_1$, \ldots, $e_n\in E$. 
Here for $s\in S$, $e\in E$, 
we let  
$$
s\cdot e= \left\{
\begin{array}{cl}
 s', & \mbox{ if } (s,e,s')\in \Tran;\\
 * , & \mbox{ if there is no } s' \mbox{ such that }(s, e, s')\in \Tran.
\end{array}
\right.
$$
This correspondence is one-to-one.
The inverse map takes any state space $(M(E,I), S_*)$ and $s_0\in S_*$ 
to an asyncronous system $(S, s_0, E, I, \Tran)$ where 
$\Tran =\{(s, e, s\cdot e)~|~ s\in S ~\&~ s\cdot e \in S\}$.
\end{lemma}

In other words, the weak asynchronous system
and hence the asynchronous
transition system can be viewed as the state space
 $(M(E,I),S_*)$ with distinguished $s_0\in S_*$.

\begin{definition}
A {\em morphism} of weak asynchronous systems $(f, \sigma):\mA\to \mA'$ 
 consists of partial maps  $f: E\pright E'$
and $\sigma: S\pright S'$ satifying the following conditions
\begin{enumerate}
\item  $\sigma(s_0)=s'_0$;
\item  for any triple $(s_1, e, s_2)\in \Tran$,
there is an alternative
$$
\left\{
\begin{array}{cl}
(\sigma(s_1), f(e), \sigma(s_2))\in \Tran', & \mbox{ if } f(e) \mbox{ is defined},\\
\sigma(s_1)=\sigma(s_2), & \mbox{ if } f(e) \mbox{ is undefined},
\end{array}
\right.
$$
\item for each pair $(e_1,e_2)\in I$ such that $f(e_1)$ and $f(e_2)$ are defined, 
the pair $(f(e_1),f(e_2))$ must belong to $I'$.
\end{enumerate}
\end{definition}

If $s_0\not=*$, $s'_0\not=*$ and $\sigma: S\to S'$ is defined on the whole $S$,
then these conditions gives a {\em morphism of asynchronous systems} 
in the sense of \cite{bed1988}.
Following \cite{bed1988} denote by $\AS$ the category of asynchronous systems.

\begin{definition}\label{homoas}
A morphism of weak asynchronous systems $(f,\sigma): \mA\to \mA'$ is 
{\em polygonal} if $(f,\sigma)$ 
defines the 
independence preserving morphism of the corresponding state spaces.
\end{definition}
Denote by $\ASb$ the category of asynchronous systems 
and polygonal morphisms.
We show that the category $\AS$ is not a subcategory of $\ASb$.

\begin{proposition}\label{critpolyg}
A morphism 
$
(\eta, \sigma): (S, s_0, E, I, \Tran)\to (S', s'_0, E', I', \Tran')
$
 in the category $\AS$ is polygonal if and only if for any  
 $s_1\in S$, $e\in E$, $s'_2\in S'$ the following implication holds
$$
(\sigma(s_1), \eta(e), s'_2) \in \Tran' ~\Rightarrow~ (\exists s_2\in S)
 (s_1, e, s_2)\in \Tran.
$$
\end{proposition}
{\sc Proof.} If  $(\eta, \sigma)$ is a polygonal morphism, then for any $s_1\in S$ 
and $e\in E$ such those  $s_1\cdot e=*$, we have
 $\sigma(s_1)\cdot \eta(e)= \sigma(s_1\cdot e)=*$.
It follows that a morphism of asynchronous systems is polygonal if and only 
if for all $s_1\in S$ and $e\in E$ the following implication holds
$s_1\cdot e = * \Rightarrow \sigma(s_1)\cdot \eta(e) = *$.
By the law of contraposition, we obtain for all $s_1\in S$ and $e\in E$ that
\begin{equation}\label{undef1} 
 (\exists s'_2 \in S')(\sigma(s_1), \eta(e), s'_2)\in \Tran' \Rightarrow 
(\exists s_2\in S)  (s_1, e, s_2) \in \Tran.
\end{equation}
Taking out from the formula   (\ref{undef1}) the variable $s'_2$  with the quantifier, we get 
$$
 (\forall s'_2 \in S')\left((\sigma(s_1), \eta(e), s'_2)\in \Tran' \Rightarrow 
(\exists s_2\in S)  (s_1, e, s_2) \in \Tran\right).
$$
Adding to the formula the quantifiers $ (\forall s_1 \in S) (\forall e \in E) $, 
we obtain the required assertion.
\hfill $\Box$

Let $\pt_*=\{p, *\}$ be a state space with the monoid 
 $M(\emptyset, \emptyset)=\{1\}$.
Associating with weak asynchronous system the morphism of state spaces 
$\pt_*\to (M(E,I), S_*)$
 defined as $p\mapsto s_0$, we obtain

\begin{proposition}\label{ptcomma}
$\ASb$ is isomorphic to the comma category
 $\pt_*/(FPCM^{\|},\Set_*)$.
\end{proposition}

For any complete category $\mA$ and object $A\in \Ob\mA$, 
the comma-category $A/ \mA$ is complete.
It follows from \ref{cpolpar} and \ref{ptcomma} the following

\begin{theorem}\label{asblim}
The category $\ASb$ is complete.
\end{theorem}

The completeness of $\AS$ is shown in \cite{bed1988}.
It follows from Propositions 
 \ref{cocpolpar} and \ref{ptcomma} the following

\begin{theorem}\label{asbcolim}
The category $\ASb$ is cocomplete.
\end{theorem}

\section{Conclusion}

There are possible applications of the results related
with building adjoint functors between the category of $ \ASb $ 
and the category of higher dimensional automata.
{\em Unlabeled semiregular higher dimensional automation} \cite{gou1995} 
is a contravariant functor from the category of cubes 
into the category $\Set$. 
Let $\Upsilon_{sr}$  be a category of unlabeled semiregular higher dimensional automata
 and natural transformations. 
By \cite[Proposition II.1.3]{gab1967} for each functor 
 from the category of cubes to the category $(FPCM^{||},\Set)$,
there exists a pair of adjoint functors between the categories
$\Upsilon_{sr}$ and $(FPCM^{||},\Set)$.
We can take the functor assigning to $n$-dimensional cube the state space $(\NN^n, h_{\NN^n})$ 
where  $h_{\NN^{op}}: \NN^{op}\to \Set$ is the contravariant functor of morphisms.
So, we get left adjoint to the composition 
$(FPCM^{||},\Set_*) \to (FPCM^{||},\Set) \to \Upsilon_{sr}$.
Taking initial point, we obtain adjoint functors between 
$\ASb$ and the category of higher dimensional automata with the initial point.

Considering the comma categories, we can compare the labelled asynchronous systems 
with labelled higher dimensional automata.

Event structures and Petri nets can be considered as asynchronous systems. 
Therefore, applications of polygonal morphisms for the study
 of Petri nets and event structures are possible.

\end{document}